# Review of Theoretical and Computational Methods for 2D Materials Exhibiting Charge Density Waves

Sugata Chowdhury [1,2], Heather M. Hill [2,3], Albert F. Rigosi [2], Patrick M. Vora [4,5], Angela R. Hight Walker [2], and Francesca Tavazza [2,*]

[1] Department of Physics and Astronomy, Howard University, Washington, DC 20059, United States; sugata.chowdhury@howard.edu
[2] National Institute of Standards and Technology (NIST), Gaithersburg, MD 20899, United States
[3] American Institute of Physics, College Park, MD 20740, United States
[4] Quantum Science and Engineering Center, George Mason University, Fairfax, VA 22030, United States
[5] Department of Physics and Astronomy, George Mason University, Fairfax, VA 22030, United States
* Correspondence: francesca.tavazza@nist.gov;

**Abstract:** Two-dimensional (2D) materials that exhibit charge density waves (CDWs) have generated many research endeavors in the hopes of employing their exotic properties for various quantum-based technologies. Early investigations surrounding CDWs were mostly focused on bulk materials. However, applications for quantum devices have since required devices to be constructed from few-layer material to fully utilize the material's properties. This field has greatly expanded over the decades, warranting a focus on the computational efforts surrounding CDWs in 2D materials. In this review, we will cover ground in the following relevant, theory-driven subtopics for $TaS_2$ and $TaSe_2$: summary of general computational techniques and methods, atomic structures, Raman modes, and effects of confinement and dimensionality. Through understanding how the computational methods have enabled incredible advancements in quantum materials, one may anticipate the ever-expanding directions available for continued pursuit as the field brings us through the 21$^{st}$ century.

**Keywords:** density functional theory; charge density waves; transition metal dichalcogenides

## 1. Introduction

Two-dimensional (2D) materials compose a rapidly evolving field within condensed matter physics, as well as related disciplines. More specifically, 2D materials that exhibit charge density waves (CDWs) have generated great interest given the relevance of their exotic properties to various quantum-based technologies [1-6]. These materials are characterized by significant in-plane bonding and relatively weak out-of-plane van der Waals interactions, the latter of which allows for the isolation of individual monolayers through mechanical exfoliation. These monolayers can then enable the fabrication of many thin layered devices capable of confirming properties and behaviors determined by computational methods.

A CDW is a quantum phenomenon that takes place in correlated systems [7-11], and the states exhibiting this phenomenon are of profound theoretical interest since such states may provide a switchable function for tuning the electrical properties of customized devices. In short, the CDW state may be described as a periodic distortion of the atomic position, along with a corresponding modulation of the electron density, both of which occur below a critical transition temperature [8]. The periodicity with which a CDW may be exhibited has two major forms. In the first form, the value takes on an integer multiple of the undistorted lattice constant (also called a commensurate CDW, or C-CDW). The second form, namely the incommensurate CDW (or IC-CDW), has an unrelated periodicity and typically has a much higher transition temperature. Once a material undergoes a phase transition at low enough temperature, the change in lattice periodicity is accompanied by a localized band splitting at the Fermi level, lowering the overall energy of the system.

Experimental and theoretical investigations into CDW phase transitions started over 50 years ago and focused mostly on a special class of materials known as transition metal dichalcogenides (TMDs). Among the efforts were those that explored the theory behind CDW formation and behavior [12-20], motion and transport properties [21-25], and several other experimental works [26-28], including neutron scattering [29], impurity pinning [30], and thermal properties [31]. Moreover, two major phases of TMDs were primarily focused on: 1T and 2H. A simple way to visualize the difference between these two phases is that the 1T phase has a planar cross-section that may be described as more rhombic than its 2H phase counterpart. During this first major wave of research efforts, work was done involving systems that exploited the nature of electrons in 2D environments [32-34]. Through the 1990s and early 2000s, the work focused dominantly on a few material families of TMDs, paving the way for a fuller theoretical understanding of the origins, characteristics, and applications of CDW states [35-46].

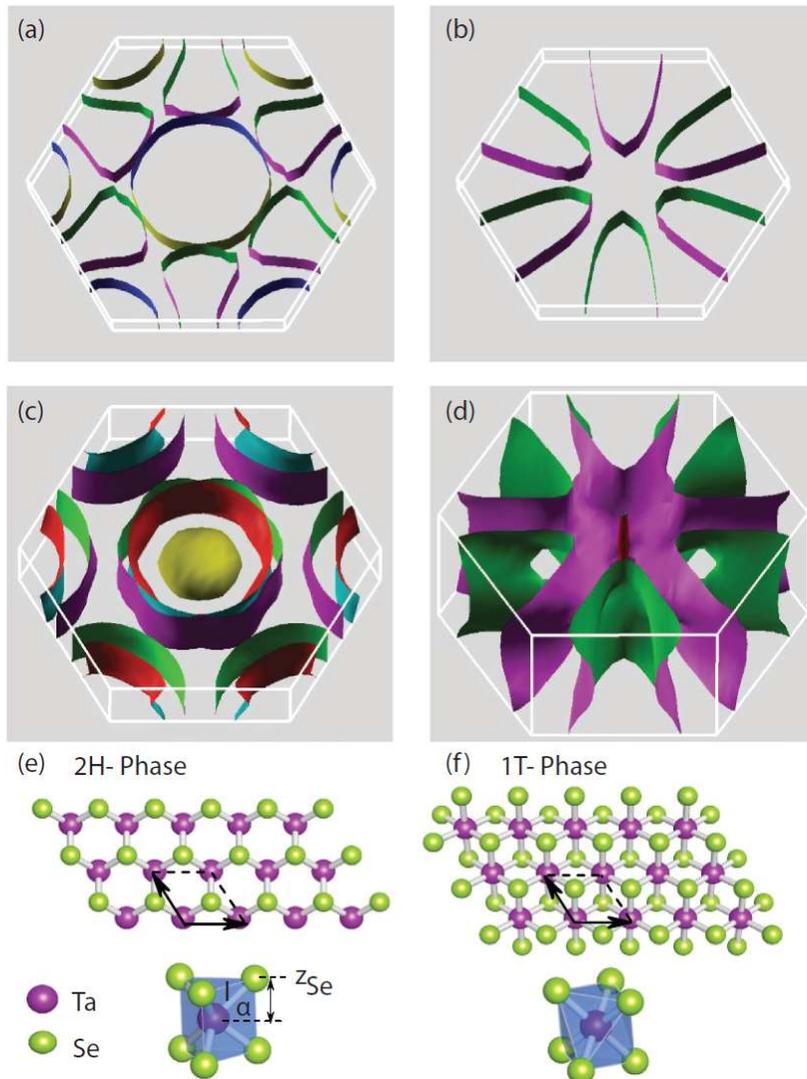

**Figure 1.** (a) Fermi surfaces are calculated for monolayer *2H*-TaSe$_2$, (b) monolayer 1T-TaSe$_2$ and (c), (d) their respective bulk counterparts with spin orbit coupling (SOC) explicitly included [82]. The crystal structures of monolayer TaSe$_2$ are shown for the (e) *2H* phase and the (f) *1T* phase. Ref. [82] is an open access article distributed under the terms of the Creative Commons CC BY license, which permits unrestricted use, distribution, and reproduction in any medium.

Within the modern era of theoretical research on CDWs in TMDs, it became evident that in order to utilize quantum devices constructed from few-layer material, a comprehensive examination of device behaviors with a variety of computational methods, like density functional theory (DFT), was crucial. These quantitative assessments of the observations made in these material systems became vital for future efforts to realize quantum information platforms based on nonequilibrium phenomena [48, 49]. It also became clear that a large variety of material systems was available for computing general physical phenomena. One nice example of the computational abilities developed was to calculate Fermi surfaces for these materials, like those shown in Figure 1. Within the last decade, there have been many theoretical and experimental results reported on the following TMD material systems: $NbSe_2$ [50-54], $NbS_2$ [55-60], $TiS_2$ [61-65], $TiSe_2$ [66-71], $VS_2$ [72-76], and $VSe_2$ [77-81].

The sizeable number of pursuits surrounding these materials renders any attempt at compiling every use or research effort impractical, so, instead, this review will cover the following relevant, theory-driven subtopics for the materials systems $TaSe_2$ and $TaS_2$: summary of general computational techniques and methods, atomic structure, Raman modes, and effects of confinement. The review will also focus primarily on theoretical work, with appropriate context and some supporting experimental efforts that ultimately validate some of the efforts described herein. Through understanding how such computational methods have enabled incredible advancements in quantum materials, one may anticipate the ever-expanding directions available for continued pursuit as this field brings us through the 21st century.

## 2. Computational Techniques and Methods

Currently, the most accurate way to computationally determine CDW Raman modes is by using density functional theory (DFT) [83, 84]. DFT can accurately predict the electronic and phononic structures of various materials and allows for the use of larger computational cells, which in turn are key to successfully carrying out a model for CDW phase transitions [46, 54, 85]. However, there are a couple of obstacles to overcome when using DFT for such purposes. The first obstacle is intrinsically related to some of the properties used in DFT: for instance, all standard DFT calculations are performed at the thermal temperature of 0 K, whereas a phenomenon like CDW phase transitions are inherently temperature-driven. A second obstacle presents itself during the use of plane-wave DFT codes. Sometimes, CDW phases have periodic cells that are incommensurate (I-CDW), thereby making any modeling of a phase transition a rather non-trivial problem if periodic boundary conditions are imposed. A third obstacle that can arise before starting any simulations is the choice of appropriate DFT parameters. The most suitable psedo-potential/exchange-correlation functional must be selected, as well as the appropriate k-point density determined by a metric of high convergence. Such choices are, obviously, not unique to the CDW modeling, but are of high importance here as the system response in very sensitive to these parameters. In the rest of this section, we will briefly review how these points were dealt with in the case of $2H-TaSe_2$ and $2H-TaS_2$.

Some DFT results presented throughout the review were carried out using the QUANTUM ESPRESSO (QE) package [86-88], and this is widely seen as a versatile tool for DFT work. There are many other approaches and details that can make a DFT approach more suitable over others depending on the material system. Other techniques can include the exchange-correlation interaction being described by local density approximations using Perdew-Wang correlation [89], norm-conserving pseudopotentials [90, 91], and van der Waals corrections [92-96]. More details on the DFT methodology can be found in additional related work [97-104].

Though there are a variety of approaches for including finite temperature in DFT simulations, like DFT-MD or finite-temperature DFT, their computational cost is significantly larger than that of exact DFT, making them difficult to use for larger simulations like those required for investigating CDW transitions. An alternative approach to avoid this problem of temperature is to model the formation of the CDW states by only changing the system's *electronic* temperature rather than concerning oneself with the thermal temperature. , In order to qualitatively assess the effect of temperature on the phonon properties of the system and the Kohn anomaly [46, 54, 105-107], a smearing factor σ, which is a parameter characteristic of the Fermi-Dirac distribution associated with a certain temperature, must be introduced. This approach of modeling real temperature effects with electronic temperature variations is computationally inexpensive and can be validated using secondary means, like comparing the lattice expansion as a function of temperature for theoretical and experimentalcases.

As an example, the case of TaSe$_2$ is shown in Ref. [108]. Here, the DFT findings reproduce the qualitative changes for both *a* and *c* lattice constants compared to the experiment and are also close to being quantitatively correct. The maximum relative change for *a* is 1.5% experimentally and 1.1% in the DFT evaluation. The thermal evolution of lattice constant *c* is not as accurate but still completely acceptable (4.1% experimentally and 2.3% computationally). Even in the more complex case of TaS$_2$, the thermal evolution trend is well-reproduced with this approach of only using the electronic temperature [109].

Though some materials, like 2H-TaSe$_2$, have a CDW phase commensurate to their room temperature phase, that is not always the case. On the other hand, 2H-TaS$_2$ is an example of a compound with an IC-CDW phase. Modeling the transition between the room temperature phase and incommensurate structure is not straightforward if periodic boundary conditions are used. To circumvent such a problem, Janner *et al*. suggested that it's possible to apply a perturbation along the *c* axis in the form of small compressive stress to computationally model incommensurability using a relatively small cell [99]. This strategy was successfully applied in Ref. [108] using a 3 x 3 x 1 cell and a -0.3% strain to keep the *c* lattice constant from changing with temperature. From this work, stressed and unstressed Raman modes were able to be predicted (see Table 1). Comparisons among the two DFT treatments and the experimental data, for all temperatures and for such a mode are shown in Figure 2.

**Table 1.** A comparison of experimental and simulated Raman modes from Refs. [99, 109], for both the stressed and unstressed cases. The X indicates a missing value in the literature.

| Mode (cm$^{-1}$) | Exp – 4K | 10K | 30K | 50K | 60K | 300K |
|---|---|---|---|---|---|---|
| $E_{2g}^2$ (shear) - **Stressed** | 26 | 30.7 | X | X | X | 32.8 |
| $^2$E$_{2g}$ (CDW) - **Stressed** | 51 | 48.8 | 46.6 | 44.6 | 42.8 | X |
| $^1$A$_{1g}$ (CDW) - **Stressed** | 79 | 76.0 | 67.6 | 63.0 | 62.6 | X |
| $E_{2g}^2$ (shear) - **Unstressed** | 26 | 34.9 | 32.3 | 33.6 | 32.0 | 31.8 |
| $^2$E$_{2g}$ (CDW) - **Unstressed** | 51 | 42.4 | 42.2 | 39.1 | 38.1 | 36.1 |
| $^1$A$_{1g}$ (CDW) - **Unstressed** | 79 | 80.4 | 80.0 | 76.3 | 75.3 | 75.1 |

The last key point that needs careful consideration when performing computational modeling of Raman spectra for these materials is the choice of exchange-correlation functional as well as pseudopotential. While this is not the place for a complete investigation of such variables, as an example of its importance, the variability of results that such a choice brings is exemplified in the case of TaSe$_2$. Table 2 displays some of these findings for lattice constants and Raman modes (for the unit cell) under various computational choices.

**Table 2.** Raman modes for TaSe$_2$, unit cell. As the average error (computed as the average of the absolute differences between experimental and DFT result for each mode) shows at a glance, using LDA with PW parametrization leads to Raman frequencies overall closer to the experimental ones.

| Raman Modes | E$_{1g}$ | $E_{2g}^1$ | $E_{2g}^2$ | A$_{1g}$ | Average Error |
|---|---|---|---|---|---|
| Exp. (Ref. [110]) | 136 | 210 | 210 | 239 | --- |
| GGA (PBE-Norm) | 147.0 | 206.3 | 206.3 | 237 | 5.1 |
| LDA (PZ-Norm) | 142.5 | 203.9 | 203.9 | 233 | 6.2 |
| GGA (PW) | 142 | 197 | 197 | 226 | 11.2 |
| LDA (PW) | 141.3 | 212.4 | 212.4 | 241.8 | 3.2 |
| GGA (PBE-Ultra) | 129.8 | 199.5 | 199.5 | 229.4 | 9.2 |
| LDA (PZ-Ultra) | 133.6 | 207.7 | 207.7 | 227.4 | 4.7 |
| GGA (PBE-PAW) | 125.1 | 198.9 | 198.9 | 230.2 | 10.5 |

As one can see from Table 2, there are two main computational approaches being exemplified. The first approach is the use of the local density approximation (LDA). Within the LDA, Perdew-Wang (PW) and Perdew-Zunger (PZ) exchange and correlation functionals were used and, in some cases, the van der Waals density functional (vdWDF) is used to capture non-local correlations. Some works used a scalar relativistic norm-conserving pseudopotential [90, 91]. The second main computational approach is the use of the general-gradient approximation (GGA), which tends to reproduce lattice constants better than the LDA. Within the GGA, submethods include the use of projected augmented wave (PAW) code or the Perdew-Burke-Ernzerhof (PBE) method. Ultimately, Raman frequencies are better modeled using the LDA, with the smallest average error (average of the absolute difference between experimental data and DFT) being found for the PW correlation and norm-conserving pseudopotential [90, 91].

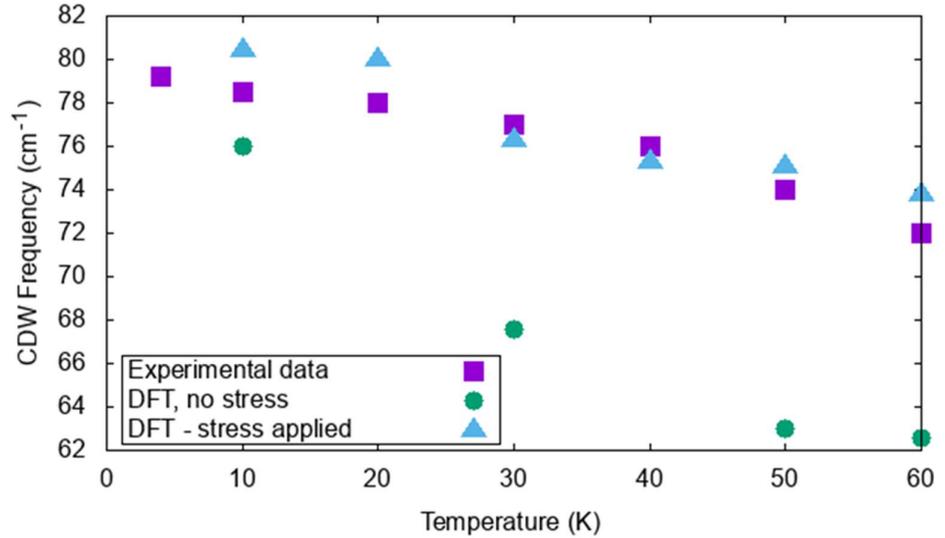

**Figure 2.** Temperature dependence of the experimentally observed CDW mode at 79 cm$^{-1}$. Including incommensurability through a small compression in the simulation cell (triangular data) leads to an excellent reproduction of the experimental data.

## 3. Computational Results in Atomic and Band Structures

The computational techniques and methods described in the previous section have all been applied in the variety of works surrounding TMDs. For the next sections, results obtained from the use of these methods will be summarized and organized by material system.

### 3.1. TaS$_2$ structure

There has been plenty of work on TaS$_2$ bulk systems, much of which uses some computational methods that have evolved into the familiar methods summarized earlier [111-113]. Regarding 2D TaS$_2$, several groups use computational methods to answer questions involving this CDW material and its structures. For instance, Lazar *et al.* analyzed the structure of thin TaS$_2$, focusing on the 1T and 2H phases [114, 115]. A commensurate charge density wave phase was derived from first principles, and both the vdW interaction and electron-electron interactions were included by using the exact exchange approximation (EXX) as well as the random phase approximation (RPA). These DFT calculations were performed using the PAW method implemented in VASP. Lazar *et al.* calculated the total energy by summing the exact exchange energy and RPA correlation energy.

Hinsche *et al.* investigated the influence of electron–phonon coupling on the electronic transport properties of both bulk and monolayer $TaS_2$. Much of the work was based on density functional perturbation theory (DFPT) and semi-classical Boltzmann transport calculations. The work reported promising room temperature mobilities and sheet conductances [116]. Though they focus on just the H-polytypes, their work implements sophisticated mathematics needed for vdW forces, namely the optB86b-vdW functional. There have been other efforts to include vdW interactions into various sample geometries [117].

Sanders *et al.* performed DFT calculations for freestanding, single layer $TaS_2$ via VASP code [118]. Others, like Kresse *et al.* [104], have also heavily relied on VASP code for similar calculations. For this picture, valence electrons were described by plane-wave basis sets with a kinetic energy threshold of 415 eV, and the exchange-correlation functional was approximated with the PBE method. To give validity to their calculations, single layer $TaS_2$ was epitaxially grown on Au(111) substrates, which yielded 2D 1*H* polymorphs. These samples were subsequently measured by angle-resolved photoemission spectroscopy (ARPES) and found to be in agreement with DFT calculations [118].

Lastly, as DFT methods improve for lower dimensional systems, the pursuit of understanding the physics in quasi-2D systems becomes warranted. Cain *et al.* performed DFT calculations on *2H*-$TaS_2$ nanoribbons to determine if any periodic features arise from CDW-type distortions [119]. Other linear chain systems had been explored in earlier theoretical work [18, 21, 25], but nanoribbons were not included in this initial effort. It turned out that the DFT calculations indicated that CDW states were supported by the nanoribbon geometries, albeit with amplitudes that were small. They also observed dramatic zigzag superstructures, though only as a result of the linear defect arrays, whose atomic structure was calculated via first principles.

### 3.2. *TaSe₂ structure*

Much like the case of $TaS_2$, work on bulk systems was plentiful in the decades preceding the experimental realization of 2D crystal exfoliation [120-126]. Once it became possible to probe lower dimensional systems, many efforts sought to fully understand CDW states in $TaSe_2$. Samankay *et al.* reported DFT calculations involving a reduced Brillouin zone of the C-CDW structure resulting from lattice reconstruction in monolayer $TaSe_2$. For their calculations, this reduced zone formed a subset of the normal Brillouin zone of undistorted $TaSe_2$. Specifically, some dozen points in the normal set were mapped to the Γ point of the reduced set. Any modes derived from this zone-folding phenomenon can result in measurable Raman spectral peaks [127].

In other work, such as that by Ge and Liu [47], the structural, electronic, and vibrational properties of both bulk and single-layer *2H*-$TaSe_2$ was investigated using first principles. They found that CDW instability remains present even after removing interlayer interactions from the performed calculations. It was also found that the Fermi surfaces of single-layer materials exhibit a strong sensitivity to spin-orbit coupling (SOC). For most of their results, the electron-core interaction was described by scalar relativistic ultrasoft pseudopotentials [47]. These calculations were performed with a plane-wave energy cutoff of 35 Ry, a 18 × 18 × 6 uniform mesh of **k** points, and the Vanderbilt-Marzari Fermi smearing method for accelerated convergence. The same group went on to perform a set of calculations based on pressure dependence, with later work including the effects of electron doping [128].

The general understanding of calculations for these CDW materials gain a solid backing when they are complimented by extensive experimental work, as seen in the case of Ryu *et al.* [129]. This work shows the characterization results of single-layer 1H-TaSe2 using ARPES, STM/STS, and DFT. A 3 × 3 CDW order was shown to persist despite numerous distinct changes in the low energy electronic structure and corresponding modifications to the topology of the Fermi surfaces. Enhanced SOC and lattice distortion were found to be a crucial component in the formation of this observed CDW order. Additional calculations include the interface structure between TaSe$_2$ and graphene, which was a used substrate. Calculating structures on other substrates, especially layered ones like epitaxially grown graphene which has an interfacial buffer layer [130, 131], has become increasingly accessible due to efforts like these. Ryu *et al.* continue by describing the intricacies of how Fermi surfaces change from these interactions.

Based on the work by Yan *et al.* [82], DFT calculations can yield DOS information for monolayer TaSe$_2$. The crystal structures of 2*H* and 1*T* phases predictably lead to electronic band dispersions that exhibit visually distinct features, as shown in Fig. 3 (a) and (c). These plots are a good example of how DFT can help with distinguishing the different phases of various 2D materials. In the case of TaSe$_2$, the distinctions in band dispersion near the Fermi level originate from the Ta atom 5*d* orbitals under the influence of the six nearest-neighbor Se atoms, which is a characteristic not present in the 1*T* case. Our last example of the usefulness of DFT calculations in understanding the atomic and electronic structures of monolayer TaSe$_2$ comes from Park *et al.* [132]. In their work, a discrepancy and expanded structural degree of freedom is understood in the case where a CDW unit cell contains thirteen Ta atoms. They go on to describe the coexistence of different CDW structures in the literature and, further, suggest the anion-centered structural model as the prime atomic structure candidate of the CDW state of monolayer 1T-TaSe$_2$. This model reproduces all observed spectroscopic features and addresses the case of SiC-caused electron doping. Overall, the heavy focus on TaSe$_2$ grants it legitimacy as a material that may provide significant progress in the development of quantum-based technologies.

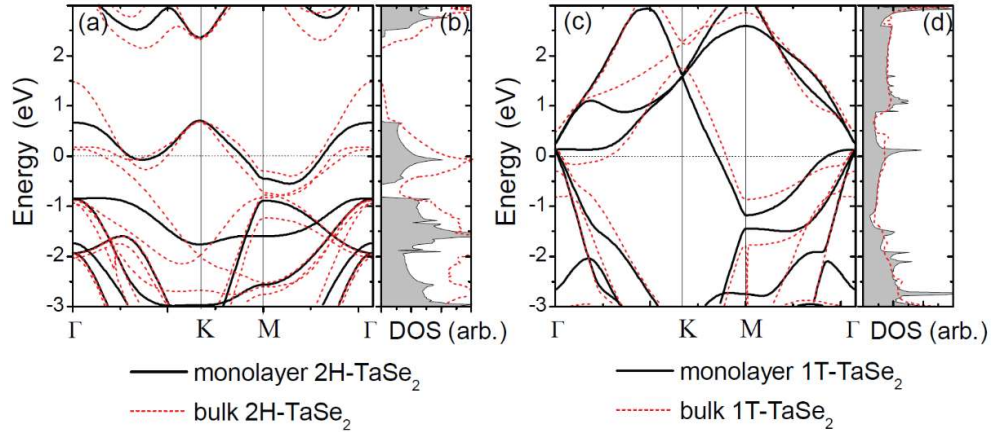

**Figure 3.** (a) Electronic band structure and (b) the DOS for monolayer *2H*-TaSe$_2$. (c) and (d) show results for monolayer 1T-TaSe$_2$. Band structures and the DOS of bulk are shown as red dashed lines. The Fermi level has been shifted to zero. Ref. [82] is an open access article distributed under the terms of the Creative Commons CC BY license, which permits unrestricted use, distribution, and reproduction in any medium.

## 4. Theory-Driven Results in Raman Modes

Raman spectroscopy has been useful as an experimental technique for distinguishing the types of modes that can arise from the unique band structures of 2D materials. Furthermore, computational methods that yield realistic models of the physics behind these materials are more than capable of generating predictions of phonon behavior, with some being verifiable with experimental data. This section focuses primarily on how computational work has enabled the verification of phonon behaviors exhibited by TaS$_2$ and TaSe$_2$.

*4.1. Phonon Behavior in TaS$_2$*

Albertini *et al.* approach the computational analysis of phonon behavior in TaS$_2$ by employing support from experimental Raman data [133]. Their first-principles calculations of the vibrational properties of 1$T$-TaS$_2$ are performed for several thicknesses in both the undistorted and C-CDW phases (high and low temperatures, respectively). Raman spectra for bulk, few-layer, and monolayer samples at low temperatures are collected and used to support the notion that low-frequency folded-back acoustic modes can act as a convenient signature of the C-CDW structure in such spectra. The observations of low-frequency modes indicates that the commensurate phase continues to be the ground state even as the material is thinned to monolayer thicknesses.

In other work, like Joshi *et al.* [109], the CDW transition is investigated with Raman spectroscopy, ARPES, and DFT. The study finds that below the CDW transition temperature, two CDW amplitude modes redshift and broaden with increasing temperature and one zone-folded mode disappears. A two-phonon mode is also observed to soften with cooling, suggesting the presence of substantial lattice distortions at temperatures as high as 250 K. These observations, backed up by DFT predictions, are further confirmed with ARPES measurements showing the persistence of a CDW energy gap above the transition temperature. The finite-temperature DFT calculations of the phonon band structure indicated an instability occurring above the CDW transition temperature, along with providing atomic displacements of the CDW amplitude modes, reproducing their temperature-dependent behavior. The authors suggest that short range CDW order exists above the transition temperature, generating new questions regarding the interplay between vibrational modes and electronic structures in layered CDW materials.

Other work revolving around phonon dispersions has been reported by Hall *et al.* [134], where the calculations of phonon dispersions were based on a combination of DFT, DFPT, and many-body perturbation theory. The phonon self-energy was calculated, and this quantity encoded the renormalization of the dispersion due to interactions of the phonons with TaS$_2$ conduction electrons. They analyze how the interplay of electronic substrate hybridization, doping, and interlayer potentials, affects the lattice dynamics and stability. This interplay shows, for example, the undoped (x = 0) and weakly hybridized ($\Gamma$ = 10 meV) case hosting a longitudinal-acoustic (LA) phonon branch exhibits a strong Kohn anomaly.

The calculations by Hall *et al.* show that hybridization is a very effective destabilizer of charge order. To verify these calculations, monolayer TaS$_2$ is grown on Au(111) and verified with ARPES and STM. The calculations show that the hybridization is energy-dependent and that the half-width at half-maximum broadening spans a range between 30 meV and 90 meV. The interplay between electron doping and the hybridization stabilized the lattice, which can be further strengthen by lattice relaxation [135].

Interesting studies on the star-of-David formations in the C-CDW state of TaS$_2$ are explored by Mijin *et al.* [136]. They used polarization-resolved, high-resolution Raman spectroscopy to investigate the consecutive CDW regimes in 1$T$-TaS$_2$ single crystals, with the support of *ab initio* calculations. Their analysis of the C-CDW spectra reveals the symmetry of the system in such a way that the preferred stacking of the star-of-David structure is trigonal or hexagonal. The spectra of the IC-CDW phase directly probed the phonon DOS due to broken translational invariance. The temperature dependence of the symmetry-resolved Raman spectra indicated that the DOS exhibited a stepwise reduction in the CDW phases, followed by a Mott transition once within the C-CDW phase, with a corresponding Mott gap to be about 170 meV to 190 meV.

*4.2. Phonon Behavior in TaSe$_2$*

When it comes to efforts pertaining to TaSe$_2$, several groups have done extensive phonon studies with computational assistance. For instance, in the case of Hill *et al.*, DFT calculations are utilized as implemented in the QUANTUM ESPRESSO package [108]. The exchange-correlation interaction was described by LDAs using the Perdew-Wang parametrization of the correlation energy [89]. Since the LDA exchange-correlation functionals yielded more reliable results when compared to experimental Raman data, as compared with the GGA [137], they became the standard for such sorts of calculations. Norm-conserving pseudopotentials were also utilized for describing the interactions between core and valence electrons.

Among their investigations was an analysis of the temperature dependence of the two-phonon mode and its origin. The required momentum for the two-phonon process came from transitions between a quasi-acoustic (QA) mode and a transverse optical (TO) mode [138-141]. The phonon band structure was computed using a unit cell at seven electronic temperatures rather than the usual $3 \times 3 \times 1$ supercell [142-146], and the results allowed for the identification of the location of the Kohn anomaly as near the Brillouin zone boundary at the M-point. The Raman spectra, shown for a range of temperatures below 130 K in Fig. 4 (a), were measured using 515-nm excitation.

To understand some of the modes, the group turned to Lee *et al.*, who predicted the appearance of two different modes in a 1D metal in a CDW state: an amplitude mode (amplitudon) and a phase mode (phason) [147]. The phase mode is a vibration representing the electron density wave in an atomic lattice which has been rearranged (whereas a phonon concerns the translation of atomic positions), and the amplitude mode is a vibration of the ions resulting in an oscillation of the intensity of the charge density and the magnitude of the CDW gap. The amplitude and phase modes are best identified by their frequency shifts, intensities, and damping rates.

It is this behavior of narrowing and intensifying of the labelled Amp 1 and Amp 2 modes with decreasing temperature that reveals them as CDW amplitude modes. As the temperature decreases, the intensity of the labelled P1 and P2 modes increases more rapidly than the intensity of the amplitude modes, but otherwise the frequencies and full width at half maximum do not vary much with temperature. This observation reveals the peaks to be phase modes, which are observed only in the C-CDW state because while in the IC-CDW state, they have only acoustic (not optical) dispersions and are therefore not Raman active. Figure 4 (c) shows the atomic vibration of the amplitude and phase modes at an electronic temperature of 10 K. The most notable vibration of the group is one of the amplitude modes, which allows the Se atoms to form a circular group vibration.

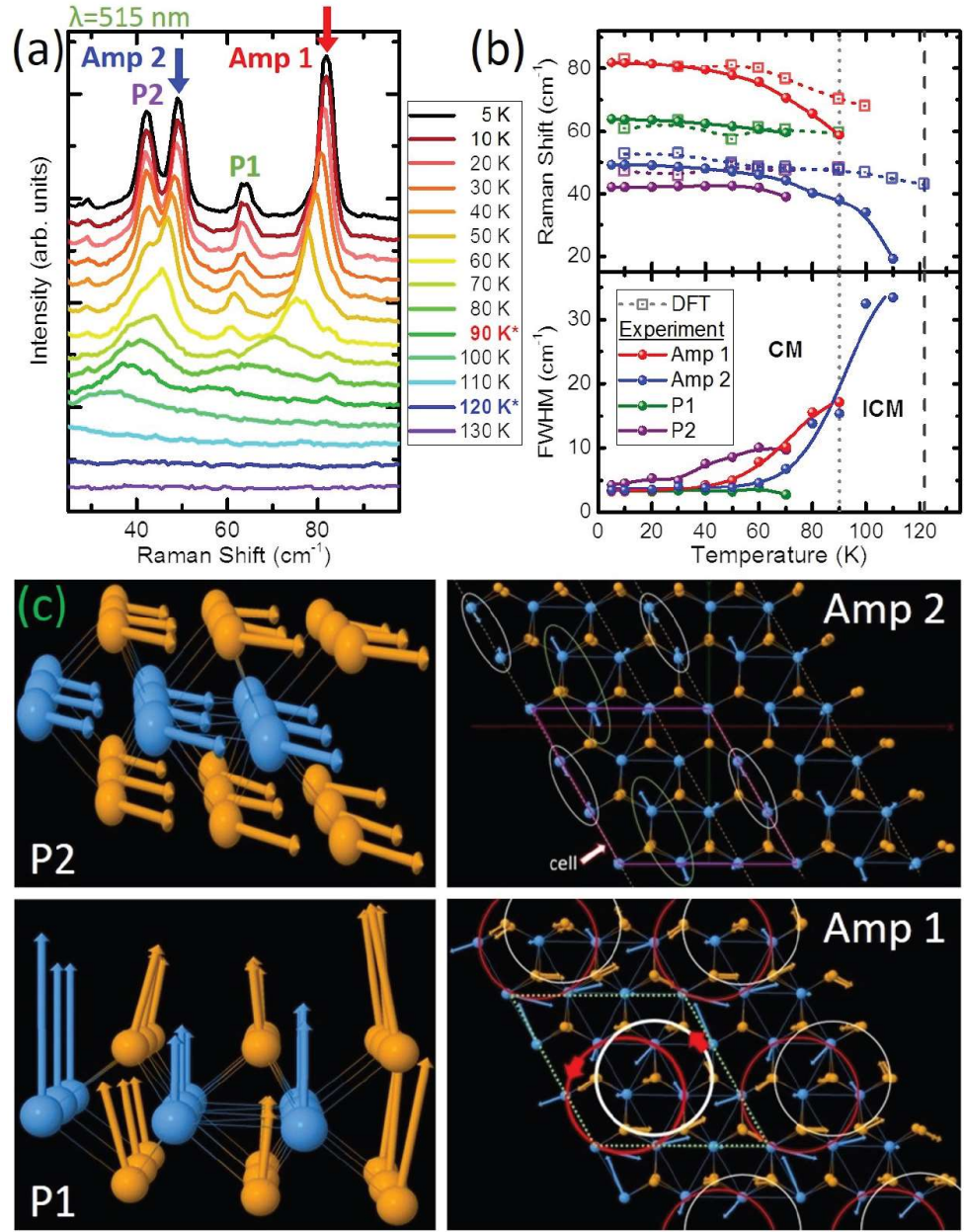

**Figure 4.** (a) Low frequency Raman spectra show four modes emerge at low temperatures: Amp 1, Amp 2, P1, and P2. Amp 1 and Amp 2 are the CDW amplitude modes whereas P1 and P2 are the phase modes. The transition temperature for the IC- (C-) phase is indicated in the legend by the blue (red) star. (b) Temperature dependence of the frequencies and full width at half maximum of Amp 1, Amp 2, P1, and P2. The DFT-calculated frequencies are plotted as a function of electronic temperature using the same temperature axis. Solid lines guide the eye. The transition temperature for the IC-CDW (C-CDW) is indicated by a dark (light) gray dashed line. (c) DFT-calculated vibrations for the phase (side view) and amplitude (top view) modes. Reprinted figure with permission from Ref. [108]. Copyright 2019 by the American Physical Society.

Chowdhury *et al.* used first-principles calculations to probe the effects of mechanical strain on the magnetic and optical properties of monolayer *2H*-TaSe$_2$ [148]. Unexpected spin behaviors were predicted, such as uniaxial-strain-induced ferromagnetism and E phonon degeneracy lifting, mainly due to the nuanced dependence of such properties on strain. For the first unexpected behavior, their calculations showed that any magnetic properties depended on the exchange within the 5$d$ orbitals of the Ta atoms. As for the effects of strain on Raman-active phonon modes, it was found that the *E* phonon degeneracy was lifted, and any electron-phonon interactions depended on the strain and its direction. These calculations revealed that CDWs weakened the magnetic properties due to symmetry breaking and Ta atom displacements. Overall, such calculations could be applicable to other CDW 2D materials.

In addition to work on $2H$-TaSe$_2$, the $1T$ phase has also had its share of academic pursuits. One example is with Wang *et al*. [149], who look at the temperature-dependent Raman spectra and phonon frequencies of 6.3 nm and 3 nm thick $1T$-TaSe$_2$. For the thinner sample, both C-CDW soft modes are observed and follow the same evolution as observed in the thicker sample. Furthermore, they remain observable up to higher temperatures compared with the thicker case. The recovery of the C-CDW state in the cooling cycle suggests that no deterioration had taken place at high temperatures (585 K). And based on the frequencies of the soft modes, the C-IC phase transition temperatures were determined to be 520 K and 570 K for thicker and thinner $1T$-TaSe$_2$ samples, respectively.

Similar work surrounding transition temperatures was performed by Samnakay *et al*. [127]. They found that the sample thickness required for inducing transition temperature changes is similar to the length of the phonon mean free path, as supported by observed modifications in the phonon spectrum of thin films as compared to bulk ones [150]. Because different CDW materials have different phase diagrams, thickness dependence trends observed for one material are not readily transferrable to another. They refer to other work, wherein a supercooled state in nm thick *$1T$*-TaS$_2$ resulted from rapid sample cooling, and this was indicative of C-CDW suppression [151]. So for TaSe$_2$, the observed thickness dependence of the phase transition temperature was able to be interpreted qualitatively within their physical model. By defining the phase transition temperature as the balance between the energy for elastic periodic lattice distortions and the energy reduced after a phase transition to the new CDW ground state, one may conclude with ease that any observed changes due to film thickness is unique to each material.

For these reasons, which extend to most 2D materials, there is a thickness dependence of the phase transition temperature that can increase or decrease depending on the material as its film thickness is modified. And though the exact mechanism that causes this phenomenon is a point of debate, the idea of confinement and dimensionality have drastic effects on CDW-exhibiting materials like TaS$_2$ and TaSe$_2$ is well established.

**5. Effects of Confinement and Dimensionality**

Since applications for quantum devices often require samples to be exfoliated from or grown into few-layer material, the heavy contributions by phenomena like vdW interactions to layer-dependent properties warrant careful treatment in calculations. This section focuses on effects of confinement and dimensionality, which can manifest frequently as comparisons between monolayer, few-layer, and bulk systems, both in computational methods and experimental data.

*5.1. Confinement and Dimensionality in TaS$_2$*

Some interesting observations can be made when a CDW-exhibiting material is forced into confinement with another adjacent material. Yamada *et al*. looked at bismuth ultrathin films on *$1T$*-TaS$_2$ and investigated electronic states with ARPES and first-principles band structure calculations [118, 152]. Though some ARPES work has been shown earlier [118], this work explores the CDW proximity effect. It was then found that the film on *$1T$*-TaS$_2$ underwent a major structural transition from (111) to (110) upon reducing the thickness, and the observation was accompanied by changes in the energy band structure. They go on to discuss the mechanisms for those observations.

Other groups have explored coexisting states, like Yang *et al* [153]. TMDs with both CDW and superconducting orders can offer ideal platforms for exploring the effects of dimensionality on correlated electronic phases. According to the authors, dimensionality has a large effect on superconductivity and CDW instabilities. They report enhancement of the superconducting transition temperature from sub-Kelvin temperatures in bulk to 3.4 K for monolayer $2H$-TaS$_2$. It is also found that the transport signature of a CDW phase transition vanishes as one approaches monolayer thicknesses. It is shown that the root cause for the enhancement stems from a reduction of the CDW amplitude, resulting in a substantial increase of the DOS at the Fermi energy. This competition between CDW and superconductivity as one approaches the 2D limit offers deep insights in our understanding of correlated electronic phases in reduced dimensions. Similar topical work has explored the relationship between CDWs and superconductivity [154-155]. Castro Neto proposed that, because of the strong variations in electron-phonon coupling as well as imperfect nesting of the Fermi surface, the CDW gap is sixfold symmetric. Ultimately, such a configuration leads to the reduction of electron-phonon scattering. Since lattice inversion symmetry is lost during the CDW phase, Dirac electrons become coupled to acoustic phonons, leading to a damping of the Dirac electrons. Lastly, these phonons drive the system to a superconducting state via a Kosterlitz-Thouless phase transition [154]. Koley *et al.* [155], on the other hand, looked to understand the competition between CDW and superconducting states by adding nonmagnetic disorder to them. Both random disorder and clustered disorder were explored using a self-consistent Bogoliubov-de-Gennes formulation. Theoretical observations of the evolution of both states were made having used a Monte Carlo method [155].

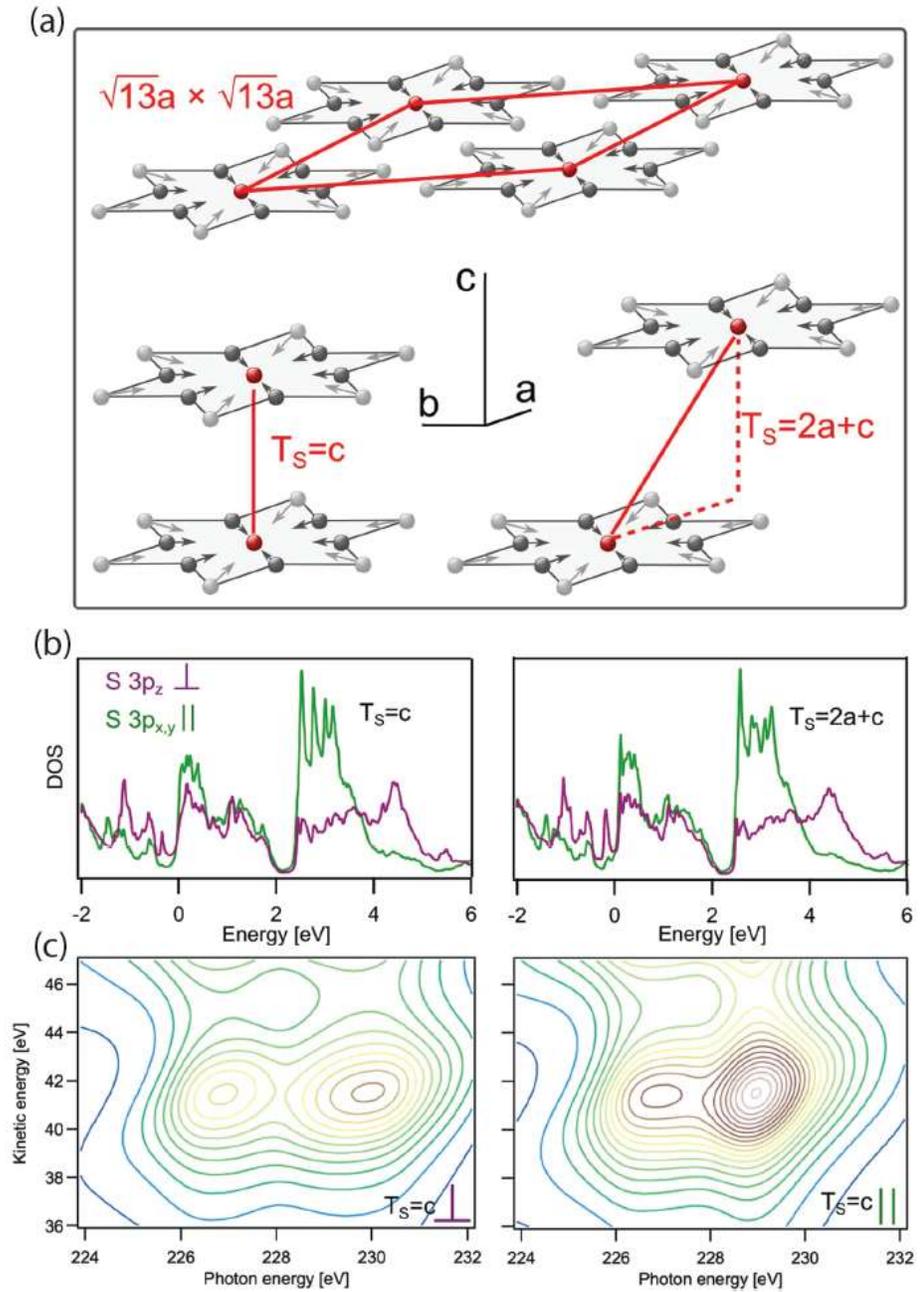

**Figure 5.** DFT calculations of the electronic structure and spectral signatures of *1T*-TaS$_2$ are shown for two C-CDW interlayer stacking arrangements. (a) The 13 *a* × 13 *a* Star-of-David reconstruction of the Ta layers is illustrated. All atoms of each star are displaced (depicted with arrows) towards the center atom. Bottom left: CDW stacking of the Ta planes with the Star-of-David centers on top of each other ($T_s$ = c). Bottom right: CDW stacking with a stacking vector having an in-plane component ($T_s$ = 2a + c). (b) DFT calculations of the PDOS of the polarized S 3*p* states with on-site S 2*s* core hole for $T_s$ = c (left) and $T_s$ = 2a + c (right). Purple shows the out-of-plane PDOS and green shows the in-plane PDOS. The Fermi level is located at 0 eV. (c) Simulations of the sulfur scattering planes based on the PDOS from (b). Intensity increases from blue to brown. Ref. [157] is an open access article distributed under the terms of the Creative Commons CC BY license, which permits unrestricted use, distribution, and reproduction in any medium.

Some works surrounding the effects of confinement or dimensionality are focused on the layer-by-layer response of the material. For instance, Ishiguro *et al*. report a layer-dependence study centered on the transition between the IC-CDW and NC-CDW and utilized electrical resistance measurements and Raman spectroscopy [156]. It was found that temperature hysteresis decreases in this transition as the number of layers decreases, and this observation is in contrast to the C-CDW and NC-CDW transition. This hysteresis trend being different for these two types of transitions was explained by the difference in the CDW superstructure along the out-of-plane direction between the C-CDW and IC-CDW phases.

Kuhn *et al*. found that even though *1T*-TaS$_2$ is a conventional 2D material, isotropic 3D charge transfer mechanisms appear in the C-CDW phase, indicating strong coupling between layers [157]. To observe and understand this oddity of dimensionality, they use a unique experimental approach and DFT to specifically probe how ultrafast charge transfer takes on a 3D or 2D character, depending on the circumstances. The experimental data are collected by what is known as the X-ray spectroscopic core-hole-clock method, which enables the experimenter to prepare in- and out-of-plane polarized sulfur 3$p$ orbital occupation (with respect to the plane of *1T*-TaS$_2$) for monitoring sub-fs wave packet delocalization. The details for some of their work are shown in Figure 5.

For this study, DFT was implemented via the Siesta code and used to compute the electronic structure of *1T*-TaS$_2$ in the C-CDW phase [158]. The calculations used the vdW-DF by Dion *et al*. [117] with the optimized exchange from the work of Klimes *et al* [159]. Ultimately, the theory shows that interlayer interactions in the C-CDW phase cause isotropic 3D charge transfer as opposed to layer stacking variations, and this mechanism for phase transitions is applicable to other CDW materials. In the case of TaS$_2$, the projected DOS is distributed over a wide energy range, reflecting the covalent bonding with neighboring atoms in the material, and observations have been made via inverse photoemission experiments [160, 161].

*5.2. Confinement and Dimensionality in TaSe$_2$*

When talking about confinement and dimensionality in TaSe$_2$, some similar conclusions may be drawn from material behaviors as one approaches the 2D limit. Yan *et al.* talks about the band dispersions for bulk *2H*-TaSe$_2$ and *1T*-TaSe$_2$ and describes what happens as the dimension is reduced from 3D to 2D [82]. Electrons in monolayer TaSe$_2$ become confined into the plane, offering a genuine 2D character to the electronic structure. However, as one could surmise from the earlier case of TaS$_2$, one may also expect that the reduced dimensionality alters the Fermi surface topology, and this alteration implies that CDW transitions are subject to dimensionality effects.

Other work, like one from Chowdhury *et al*. [162, 163], looked to learn more about how these dimensionality and confinement effects contribute to structural formation, and more specifically, to the formation of CDW states. Though various efforts have been made, many making observations or calculations that directly involve the effect of vdW interactions and reduced dimensionality on the CDW phase [164, 165], the level of detail was still lacking as far as how exactly a CDW state formed as a function of layer number. In summary, 2D systems formed triangular structures and bulk systems formed striped structures while in the C-CDW phase. The 2D case reflected the lack of presence of long-range order along the z-axis normally attributable to vdW interactions. So naturally, the bulk case can yield a striped structure with decreasing temperature because of the presence of long-range order.

Chowdhury *et al.* expand the work to address the natural question of whether or not the presence of interacting, neighboring layers affects these ground state structural formations and whether there is a coherent layer-dependence that can link the two extreme cases of 1L and bulk. They used DFT to investigate the 2L, 3L, 4L, 5L, and 6L cases, with all cases starting from the same type of super-cells as built for 1L and bulk, and with all cases being examined as the electronic temperature was reduced from 300 K to 10 K.

Sometimes confined systems, like these 2D CDW materials, are subjected to conditions that may bring about a 3D-like nature. Such was the case in Ji *et al* [166], who reported on the optical suppression of the C-CDW over a sub-ps timescale. The group claims that the short timescales involved in establishing the stacking order implies that the nucleation of the IC-CDW phase is inherently of a 3D nature. The energy of the optically excited electronic state dissipates energy into modes representing periodic lattice distortions through strong electron-phonon coupling. Some of their theoretical approach is depicted in Figure 6. Overall, their results highlight the importance of considering the 3D nature of CDWs for TaSe2 and related materials. Corresponding band structure and joint DOS calculations were performed using the ELK code [167].

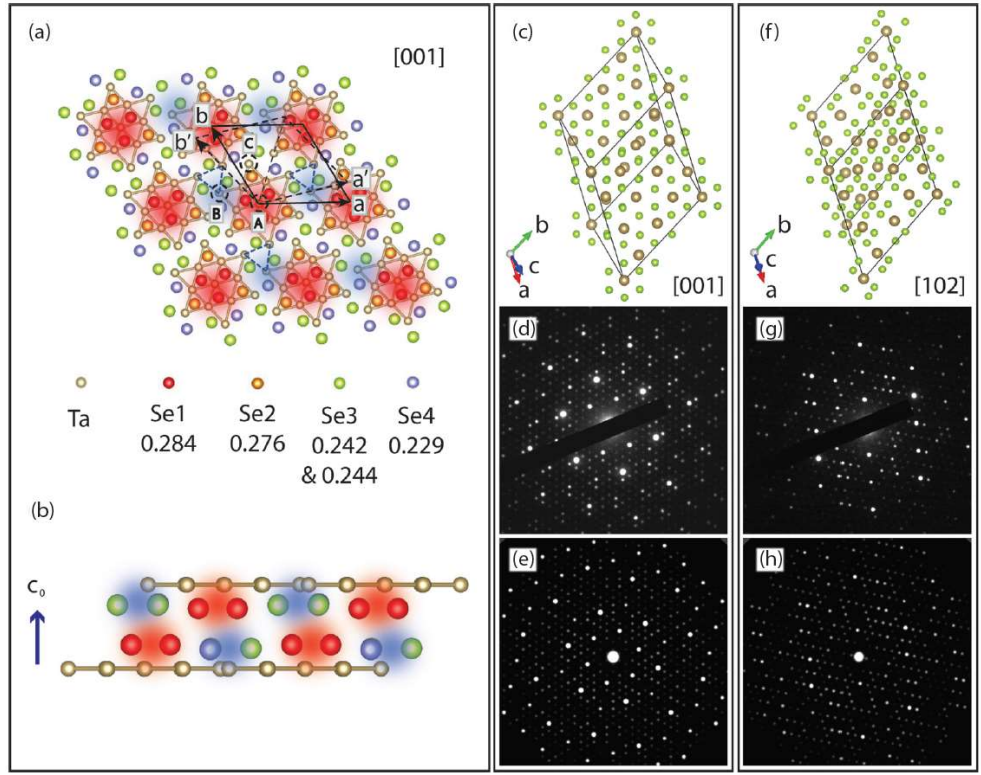

**Figure 6.** (a) Illustration of superstructure and stacking arrangement in $1T$-TaSe2. The periodic lattice displacements form a 13-Ta atom Star-of-David cluster (solid bonds). Blue dashed triangles indicate the threefold symmetric stacking displacements. Se atoms are color coded according to their $c$-axis coordinate (in fractions of $c$) as listed in the lower part of (a). The Se atoms at the center of a Star-of-David cluster are at highest (most protruding, red-shaded areas) positions, whereas other Se atoms at the lowest positions at shown as blue-shaded areas. (b) Stacking arrangement along the $c_0$ direction for two adjacent layers of the C-CDW phase. Projections of the C-CDW phase at (c) [001] and (f) [102] zone axis are shown with the corresponding experimental diffraction patterns in (d) and (g), respectively. The simulated diffraction patterns are shown along [001] (e) and [102] (h). Ref. [166] is an open access article distributed under the terms of the Creative Commons CC BY license, which permits unrestricted use, distribution, and reproduction in any medium.

Lastly, a general framework for dimensionality-dependent CDWs in both TaS2 and TaSe2 was proposed by Lin *et al* [168]. They focus on $2H$-MX2 (M = Nb, Ta and X = S, Se). One of the key takeaways in the context of dimensionality comes from their description of how ionic charge transfer, electron–phonon coupling, and the spatial extension of the electronic wave functions contribute to CDW ordering and its thickness dependence. These three parameters were used to create a unified phase diagram describing instabilities in low-dimensional limits. This diagram is illustrated in Fig. 7 (c). As can be seen, the interplay between these three parameters is expected to form an abyss-like shape, deepening at regions with low ionic charge transfer and electron–phonon coupling as well as high spatial extensions. Above the surface of the resulting terrain is a parameter-space within which the CDW is forbidden. Below this space, the CDW is allowed to emerge and does so for some of the example materials.

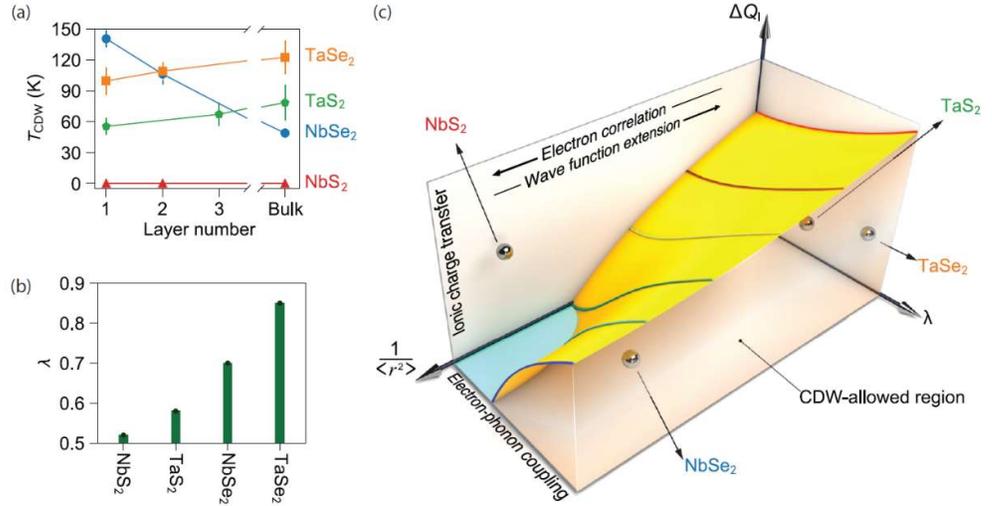

**Figure 7.** (a) Thickness dependence of the CDW transition temperature in *2H*-MX$_2$. Error bars are standard deviations obtained from the least-squares fits to the temperature-dependent amplitude mode intensity. (b) The calculated values of electron–phonon coupling constant ($\lambda$) for monolayer *1H*-MX$_2$. (c) Schematic illustration of the possible phase diagram describing the CDW response in a layered material in terms of ionic charge transfer ($\Delta Q_I$), electron–phonon coupling constant ($\lambda$), and the spatial extension of electronic wave functions ($1/\langle r^2 \rangle$). Ref. [168] is an open access article distributed under the terms of the Creative Commons CC BY license, which permits unrestricted use, distribution, and reproduction in any medium.

## 6. Future Outlook and Conclusion

Some key takeaways in this work show that even materials exhibiting similar band structures like TaSe$_2$ and TaS$_2$ have considerable differences. For instance, TaS$_2$ has a more pronounced non-commensurate transition between room temperature and its CDW phase, whereas TaSe$_2$ does not. Furthermore, inspecting the phase diagrams of the many works presented herein will reveal that TaS$_2$ will form clusters of star-of-David regions as the CDW phase is warmed, and such clusters do not form with TaSe$_2$. Rather, curvilinear regions resembling a grain boundary appear. In terms of vibrational properties, the IC-CDW phase more pronounced in TaSe$_2$ more greatly affects Raman mode osbervations between room temperature and temperatures corresponding to the C-CDW phase transition. In terms of dimensionality, a key point is that both materials differ significantly on their calculated electron-phonon coupling constant [168], and since such coupling is tied to superconductivity, it follows that observations of the latter would also be substantially altered.

An interesting direction to take would be to expand on the exploration of having non-uniformly doped areas with varying electron-phonon coupling by utilizing alternative contacting and large-scale methods for creating regions of assorted doping [52, 169-172]. It has become clear that computational methods like DFT have resulted in wide-ranging examinations of various material systems exhibiting CDWs, and such systems have the potential to become building blocks for multifaceted quantum information platforms. It is also clear that many of these material systems are slowly becoming more available for computing applications. A future roadmap for these materials and applications involves a successful identification and validation of certain TMDs that will most likely support such applications. A close coupling between theoretical and experimental efforts will ensure that existing data on these material platforms are used to iteratively guide the construction of appropriate models for newer devices, greatly accelerating device optimization.

Likewise, subsequent experimental verification of newer models should further advance the fabrication of an improved generation of TMD devices for experimental characterization, closing the feedback loop between theory and experiment. This cycling will greatly accelerate the fundamental understanding of CDWs in these matierals, as well as the limitations one may expect in their applications. Future computing demands may involve more complex systems or defect behavior, and such modeling will eventually require the development of new computational workflows and substantial computational resources.

It must also be understood that these endeavors contain inherent risk in terms of applicability, so risks must be considered and mitigated in order to reach these promising technical goals. Because the space of all TMD materials is rather large, some candidate materials may go unexplored or be assessed too soon as not viable. Additionally, the computational methodology may be need to be adequately improved to most accurately describe the key interactions taking place in these layered materials physics. Regardless of the limitations, the ongoing interplay between theoretical and experimental efforts will be vital for continued success.

This field has greatly expanded over the decades, and much has been learned on the computational efforts surrounding CDWs in 2D materials, along with some important experiments that helped validate those computational methods. We have covered several relevant, theory-driven subtopics for $TaS_2$ and $TaSe_2$ including the general computational techniques and methods, atomic structures, Raman modes, and effects of confinement and dimensionality. Through understanding how the computational methods have enabled incredible advancements in quantum materials, one may anticipate the ever-expanding directions available for continued pursuit as the field brings us into the next decade.

**Author Contributions:** Conceptualization, methodology, software, S.C. and F.T.; validation, A.F.R., P.M.V., H.M.H. and A.R.H.; formal analysis, investigation, S.C., A.F.R., P.M.V. and H.M.H.; resources, A.R.H. and F.T.; data curation, writing—original draft preparation, S.C. and A.F.R.; writing—review and editing, S.C., A.F.R., P.M.V., H.M.H., A.R.H. and F.T.; visualization, supervision, project administration, funding acquisition, A.R.H. and F.T. All authors have read and agreed to the published version of the manuscript.

**Funding:** The APC was funded by the National Institute of Standards and Technology.

**Acknowledgments:** Commercial equipment, instruments, software, and other materials are identified in this paper in order to specify the experimental procedure adequately. Such identification is not intended to imply recommendation or endorsement by the National Institute of Standards and Technology or the United States government, nor is it intended to imply that the materials or equipment identified are necessarily the best available for the purpose. Work presented herein was performed, for a subset of the authors, as part of their official duties for the United States Government. Funding is hence appropriated by the United States Congress directly.

**Conflicts of Interest:** The authors declare no conflict of interest. The funders had no role in the design of the study; in the collection, analyses, or interpretation of data; in the writing of the manuscript, or in the decision to publish the results.